\documentclass[12pt]{article}
\usepackage{epsfig}

\begin{document}

\title{{\bf
    Thermodynamic Properties of a Solid \\
    Exhibiting the  Energy Spectrum given by the Logistic Map}}
    
\author{ 
    E. M. F. Curado\footnote{e-mail: eme@cbpf.br; also at: ICCMP and Dept. de 
    F\'{\i}sica, UnB, Bras\'{\i}lia, Brazil} and M. A. 
Rego-Monteiro  
    \footnote{e-mail: regomont@cbpf.br}   \\
    Centro Brasileiro de Pesquisas F\'\i sicas, \\ 
    Rua Xavier Sigaud 150, 22290-180 - Rio de Janeiro, RJ, Brazil}

\date{\today}    

\maketitle
 
\begin{abstract}
         
\indent 
We show that the infinite-dimensional representation of the recently 
introduced Logistic algebra can be interpreted as a non-trivial 
generalization 	
of the Heisenberg or oscillator algebra.  This allow us to construct a 
quantum Hamiltonian having the energy spectrum given by the 
logistic map. 
We analyze the Hamiltonian 
of a solid whose collective modes of vibration are described 
by this generalized oscillator and compute the thermodynamic 
properties of the model in  the two-cycle 
and $r = 3.6785$ chaotic region of the logistic map. 
	
\end{abstract}

\vspace{3cm}

\begin{tabbing}

\=xxxxxxxxxxxxxxxxxx\= \kill

\>{\bf Keywords:} \> Algebras, Statistical Models; Chaos. \\

\>{\bf PACS Numbers:} \> 02.10.Gd, 05.30.-d, 63.20.-e

\end{tabbing}

\newpage

\section{Introduction}\label{sec.intro}

In the last years, complex systems have attracted a lot of 
attention.  In 
particular, there is an intrinsic theoretical interest in 
constructing a 
Hamiltonian system having an energy spectrum   
that is or quasi-periodic or self-similar, and/or chaotic \cite{nakamura}.     
Enhancing the interest in describing 
such Hamiltonian system is the fact that some models on quasicrystals 
have a quasi-periodic or fractal energy spectrum 
\cite{review,ct,celia,murilo,hugo,coreanos,luciano}. 
On the other hand, one paradigmatic example of map that 
exhibits some of these features is the 
logistic map.  As is well-known, this map describes at the Feigenbaum 
point an example of a fractal system and appearing after this point a chaotic 
region with chaotic bands and self-similar patterns \cite{livros}. 

Recently, it was developed a three-generator algebra, called Logistic 
algebra \cite{marco}, where the eigenvalue of one generator is given 
by the logistic 
map .  We show that the infinite-dimensional representation of 
this algebra can be interpreted as a non-trivial generalization of 
the Heisenberg  or oscillator algebra and call the associated 
oscillators, 
logistic oscillators.  

We use these logistic oscillators to construct a quantum Hamiltonian, 
which is a generalization of the quantum harmonic oscillator, that 
has the energy spectrum described by the logistic map.  We apply 
these ideas 
to construct a Hamiltonian describing quasi-particle vibrations of a 
solid 
with N atoms where each quasi-particle oscillates as a logistic 
oscillator.

In section 2, we discuss the logistic algebra and its interpretation 
as 
generalized Heisenberg algebra, in section 3 we construct a model for 
a 
solid where the collective modes of motion are described as logistic 
oscillators and compute the thermodynamic functions of the model in  
the two-cycle 
and $r = 3.6785$ chaotic region of the logistic map 
$x_{n+1} = r x_{n} (1 - x_{n})$ .  Section IV is devoted to our 
conclusions. 

\section{Algebraic origin of the model}\label{sec.algebra}

\indent 
The model we are going to discuss in sections 3 and 4 has its origin 
in 
an algebraic structure called Logistic algebra \cite{marco}.  In this 
section we present the 
Logistic algebra and show that this algebra can be interpreted as a 
non-trivial extension of Heisenberg algebra.

Let us consider the algebra generated by  $J_{0}$ , $J_{\pm}$ , 
described 
by the relations \cite{marco}

\begin{eqnarray}
    J_{i} J_{+} & = & J_{+} J_{i+1}  \hspace{0.3cm} i = 0,1,2,\ldots 
    \hspace{0.3cm} ,
    \label{eq:j+}  \\
     J_{-} J_{i} & = & J_{i+1} J_{-} \hspace{0.3cm}  ,
    \label{eq:j-}  \\
     J_{+} J_{-} - J_{-} J_{+} & = & -a(J_{0}-J_{1}) \hspace{0.3cm}  ,
    \label{eq:j01}
\end{eqnarray}
where $ J_{-} = J_{+}^{\dag}$ , $ J_{i}^{\dag} = J_{i}$ and  ``$a$'' 
is a 
real constant.  Moreover, 

\begin{equation}
    J_{i+1} = r J_{i} (1 - J_{i}) \hspace{0.3cm} i=0,1,2,\ldots 
    \hspace{0.3cm} ,
    \label{eq:logistic}
\end{equation} 
with  $0 \leq r \leq 4$ .  

The hermitian operator $J_{0}$ can be diagonalized. Consider the 
state 
$|0>$ with the lowest \footnote{Due to the use of the logistic map, 
depending on the values of $r$ and $\alpha_{0}$ considered, $|0>$ can 
be the state with highest weight. We emphasize in this paper the case 
where $|0>$ is a lowest weight vector since it is the situation 
considered in the following sections.} eigenvalue of $J_{0}$ 

\begin{equation}
    J_{0} |0> = \alpha_{0} |0>   \hspace{0.3cm}  . 
    \label{eq:vacuo}
\end{equation} 

\noindent 
Note that, for each value of $\alpha_0$ we have a different vacuum 
that for simplicity all of them will be denoted by $|0>$. 
We choose $0 \leq \alpha_{0} \leq 1$ because with this condition 
all the future iteration will remain in this interval and the connection 
with the chaotic concepts is straightforward. Also, the allowed 
values of $\alpha_{0}$ depend on $r$ and $a$.
Since, by hypothesis, $ \alpha_{0} $ is the lowest $J_{0}$ 
eigenvalue we must have, 

\begin{equation}
    J_{-} |0> = 0  \hspace{0.3cm} .
    \label{eq:zero}
\end{equation}

Following the usual steps for constructing (now from lower to higher 
eigenvalues) SU(2) algebra representations \cite{su2}, using the algebraic 
relations exhibited in eqs. (\ref{eq:j+},\ref{eq:j-},\ref{eq:j01}) and 
taking into 
account eqs. (\ref{eq:vacuo},\ref{eq:zero}) we obtain: 

\begin{eqnarray}
    J_{0} |m> & = & \alpha_{m} |m> \hspace{0.3cm} ,
    \label{eq:j0m}  \\
   J_{+} |m> & = & N_{m} |m+1> \hspace{0.3cm} ,
    \label{eq:j+m}  \\
    J_{-} |m+1> & = & N_{m} |m> \hspace{0.3cm} ,
    \label{eq:j-m1}
\end{eqnarray}

where \footnote{Note that if we put $m=-1$ in the eq. (\ref{eq:j-m1}) we 
obtain consistently eq. (\ref{eq:zero}).}

\begin{equation}
    N_{m} = \sqrt{a(\alpha_{0} - \alpha_{m+1})}
    \label{eq:nm} \hspace{0.5cm} ,
\end{equation}

\noindent 
and 

\begin{equation}
    \alpha_{m+1} = r \alpha_{m} (1-\alpha_{m}) \hspace{0.4cm} .
    \label{eq:logistic2}
\end{equation} 

\noindent 
Note that the states $|m>$ , $m \geq 1$ , are defined by the 
application of $J_{+}$ on $|m-1>$ .
Moreover, from eqs. (\ref{eq:j0m} - \ref{eq:j-m1}) we easily obtain a 
general expression for $|m>$,  

\begin{equation}
    |m> = \frac{1}{\prod_{i=0}^{m-1}N_{i}} (J_{+})^m |0> \hspace{0.3cm}.
    \label{eq:mgeral}
\end{equation}

Of course, since the eigenvalues of $J_{0}$ are given by the logistic 
map (eq. (\ref{eq:logistic2})), their values as $m$ increases can 
have 
an irregular behavior depending on the values of $r$ and 
$\alpha_{0}$ , 
and the dimension of the representation.  Note that, unlike $SU(2)$ 
algebra 
where the states obtained by the application of $J_{+}$ always have 
higher $J_{0}$ eigenvalues, for the logistic algebra this depends on 
what values of $r$ and $\alpha_{0}$ we consider and the level of 
iterations 
(the number $m$ of $|m>$) we are.  For instance, for $r=3$ and 
$\alpha_{m} = 0.5$ we have $\alpha_{m+1} = 0.75$ , i.e., 
$J_{+}$ rises 
the $J_{0}$ eigenvalue of $|m>$ .  On the other hand, for $r=1.5$ and 
$\alpha_{m} = 0.5$ we have $\alpha_{m+1} = 0.375$ and 
in this case $J_{+}$ 
lowers the $J_{0}$ eigenvalue of $|m>$ .  Moreover, due to the 
non-regular behavior of the logistic map, it 
may 
happen for $J_{+}$ that even having started as lowering  
the $J_{0}$ 
eigenvalue of $|m>$ it rises the $J_{0}$ eigenvalue of 
$J_{+}| m>$ for 
a given level $m$ of iteration of the logistic map.  For instance, 
for 
$r=2.75$ and $\alpha_{m} = 0.9$ we have 
$\alpha_{m+1} = 0.247$ and 
$\alpha_{m+2} = 0.5122$ . 

Let us now consider the operator

\begin{equation}
    C = J_{+} J_{-} + a J_{0} = J_{-} J_{+} + a J_{1} \hspace{0.3cm} .
    \label{eq:casimir}
\end{equation}

\noindent 
Using the algebraic relations (eqs. 
(\ref{eq:j+},\ref{eq:j-},\ref{eq:j01})) it 
is easy to see that 

\begin{equation}
    [C, J_{0}] = [C, J_{\pm}] = 0 \hspace{0.3cm} ,
    \label{eq:cascomut}
\end{equation}
i.e., $C$ is the Casimir operator of the algebra.  In fact, we arrive 
easily 
at 

\begin{equation}
    C | m> = c_{0} | m> \hspace{0.3cm} ,
    \label{eq:caseigen}
\end{equation}
with $c_{0} = a \alpha_{0}$ independent of $m$ .

With respect to matrix representations of the Logistic algebra there 
are 
finite-dimensional matrix representations corresponding to the 
n-cycle 
solutions of the logistic map and infinite-dimensional ones relative 
to the 
n-cycle and to the chaotic regime of the logistic map.  Here we 
present 
some examples: 

1.   Two-Dimensional Representations

\begin{equation}
J_{0}= \left(     
\begin{array}
[c]{cc}
\alpha_{0}^À & 0 \\
0 & \alpha_{1}^À \\
\end{array}   \right) \hspace{0.2cm} , \hspace{0.4cm}
J_{+}= \left(     
\begin{array}
[c]{cc}
0 & 0 \\
N_{0}^À & 0 \\
\end{array}   \right) \hspace{0.2cm} , \hspace{0.4cm}
J_{-} = J_{+}^{\dag} \hspace{0.4cm} .
\end{equation}
The allowed values of $r$ and $\alpha_{0}$ are determined by the 
equation 
$N_{1}^2 = 0$ such that $N_{0}^2 \neq 0$ .  There are two non-trivial 
solutions, 

\begin{equation}
    \alpha_{0}^{\pm} = \frac{r + 1 \pm \sqrt{r^2 - 2r -3}}{2r} 
    \label{eq:avciclo2}
\end{equation} 
The solution $\alpha_{0}^{+}$ gives $\alpha_{0}^{+} > \alpha_{1}^{+}$ 
implying $a>0$ , while $\alpha_{0}^{-} < \alpha_{1}^{-}$ gives 
$a<0$.  For 
both cases $r \geq 3$. We will use this solution in the next section.

2. Three-Dimensional Representations

\begin{equation}
    J_{0}= \left(  
    \begin{array}{ccc}
        \alpha_{0} & 0 & 0  \\
        0 & \alpha_{1} & 0  \\
        0 & 0 & \alpha_{2} 
    \end{array} \right) \hspace{0.2cm} , \hspace{0.4cm}
       J_{+}= \left(  
    \begin{array}{ccc}
        0 & 0 & 0  \\
        N_{0} & 0 & 0  \\
        0 & N_{1} & 0 
    \end{array} \right) \hspace{0.2cm} , \hspace{0.4cm}
    J_{-} = J_{+}^{\dag} \hspace{0.2cm} .
    \label{eq:ciclo3}
\end{equation}
The allowed values of $r$ and $\alpha_{0}$ are computed from 
$N_{2}=0$ , 
$N_{0}, N_{1} \neq 0$ . 

\begin{figure}[hbt]
\centerline{\epsfig{file=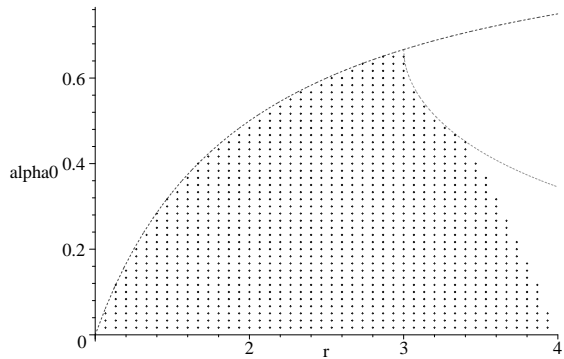,width=8cm,angle=-90}} 
\caption{{\small Region of allowed values for $\alpha_{0} $ and $r$. } } 
\label{alfar0} 
\end{figure}

3. Infinite-Dimensional Representations

\begin{equation}
    J_{0}= \left(  
    \begin{array}{ccccc}
        \alpha_{0} & 0 & 0 & 0 & \ldots  \\
        0 & \alpha_{1} & 0 & 0 & \ldots  \\
        0 & 0 & \alpha_{2} & 0 & \ldots  \\
        0 & 0 & 0 & \alpha_{3} & \ldots  \\
        \vdots & \vdots & \vdots & \vdots & \ddots 
    \end{array}
    \right)  , \hspace{0.1cm}
       J_{+}= \left(  
    \begin{array}{ccccc}
        0 & 0 & 0 & 0 & \ldots  \\
        N_{0} & 0 & 0 & 0 & \ldots  \\
        0 & N_{1} & 0 & 0 & \ldots  \\
        0 & 0 & N_{2}^À & 0 & \ldots  \\
        \vdots & \vdots & \vdots & \vdots & \ddots 
    \end{array}
    \right) , \hspace{0.1cm}
    J_{-} = J_{+}^{\dag} \hspace{0.1cm} .
    \label{eq:cicloinf}
\end{equation}
The allowed values of $r$ and $\alpha_{0}$ can be computed for 
instance for 
$a<0$ from $N_{m}^2 = |a|(\alpha_{m+1} - \alpha_{0})$ by imposing 
$\alpha_{m} > \alpha_{0}$ for all values of $m \geq 1$ .  In fig. 
\ref{alfar0} 
we show a half-leaf region with the allowed values of $r$ and 
$\alpha_{0}$ 
satisfying the above requirements.  These solutions will be used in 
the 
following section.  

Let us show now an interesting connection of this algebra with the 
Heisenberg algebra. 
The Heisenberg algebra is generated by the elements  $A$ and $A^{+}$ 
satisfying the relations  

\begin{eqnarray}
    A A^{\dag} - A^{\dag} A  & = & 1 \hspace{0.3cm} ,
    \label{eq:aa+}  \\
    N A^{\dag} - A^{\dag} N & =  & A^{\dag}  \hspace{0.3cm} ,
    \label{eq:na+}
\end{eqnarray}
with $N = A^{\dag} A$ is the number operator.  
Note that eqs. (\ref{eq:j+}, \ref{eq:j-}), for $i=0$ can be seen as 
defining 
equations for $J_{1}$.  The Heisenberg algebra comes naturally if we 
put 
in eqs. (\ref{eq:j+},\ref{eq:j-}, \ref{eq:j01}) $J_{-} \equiv A$, 
$J_{+} \equiv A^{\dag}$, $J_{0} \equiv N$, $J_{1} = J_{0} + 1$ and 
$a=-1$ .  It can be easily verified that we do not have in this case 
finite dimensional representations and the Casimir operator is 
identically null.

In summary, Heisenberg algebra is the special case of the defining 
relations 
given by eqs. (\ref{eq:j+},\ref{eq:j-}, \ref{eq:j01}) where instead 
of 
taking the relation given by eq. (\ref{eq:logistic}) we consider the 
simpler one $J_{1} = J_{0} + 1$ .  In other words, the Logistic 
algebra can 
be interpreted as an extension of the Heisenberg algebra where 
instead of the 
simple iteration $J_{i+1} = J_{i} + 1$ we take the logistic map for 
$J_{i+1}$ 
as in eq. (\ref{eq:logistic}) .  Clearly, it is also 
possible here to consider other maps; this study is under 
progress .

Of course, since the Heisenberg algebra is a master algebra in 
physics, it is 
a natural step to investigate the possible consequences of the 
logistic 
generalization, explained before, in physical problems.  In the 
following 
sections we apply this generalized Heisenberg algebra to a collective 
modes 
of motion of N atoms.

\section{Model and thermodynamic properties}
\label{sec.termodinamica}

\indent

Let us consider the Hamiltonian of a quantum system of 
quasi-particles 
described by $N$ independent, localized, ``oscillators'' of the form 

\begin{equation}
    H = \sum_{q=1}^{N} \epsilon_{q} J_{0}^{^q} \hspace{0.3cm} ,
    \label{eq:hamiltoniano}
\end{equation}
where $\{J_{0}^{q}\}$ is a collection of N independent oscillators, 
each of them 
described by the algebra (1-3), and $\epsilon_{q}$ is 
a parameter associated to the energy of the $q$-th 
oscillator.  
We are then considering independent collective 
excitations with a non-trivial spectrum specified by 
the eigenvalues of $J_{0}^q$.
For the solution $J_{-} = A$ , $J_{+} = A^{\dag}$ , 
$J_{0} = N$ , 
$J_{i+1} = J_{i}+1$ and $a = -1$ of the algebra (1-3), the 
Hamiltonian 
(\ref{eq:hamiltoniano}) describes the well-known system of N 
independent, 
localized, harmonic oscillators.  On the other hand, by considering 
the logistic 
generalization, eq. (\ref{eq:hamiltoniano}) 
becomes the Hamiltonian of a system of quasi-particles described by N 
independent, localized, logistic oscillators.  We interpret 
$J_{-}^{q}$, 
$J_{+}^{q}$ and $J_{0}^{q}$ as annihilation, creation and
generalized number 
operator, respectively, of the $q$-th oscillator. Note that the 
energy of the $q$-th oscillation-mode in a state $|m>$ 
is given by the product of $\epsilon_{q}$ times the eigenvalue of 
$\{J_{0}^{q}\}$ applied on that state.  The eigenvalue 
$\alpha_{m}^{q}$  
indicates that the $q$-th oscillation-mode is in the state $|m>$. 
We are adopting this model due to its simplicity but these 
logistic oscillators 
could also be used in more complicated models as for example in 
disordered systems.

The partition function of the model (\ref{eq:hamiltoniano}) 
 
\begin{equation}
    Z = Tr \exp{(- \beta H)} \hspace{0.3cm} ,
    \label{eq:particao}
\end{equation}
with $\beta = (k_{B} T)^{-1}$ and $k_{B}$ the Boltzmann constant, 
factorizes 
into a product of single particle partition functions, 

\begin{eqnarray}
    Z & = & \prod_{q} Z_{q} \hspace{0.3cm} ,
    \label{eq:z1}  \\
    Z_{q} & = & \sum_{m=0}^{\infty} \exp{(- \beta \epsilon_{q} 
\alpha_{m}^{q})} 
    \hspace{0.3cm} ,
    \label{eq:zl}
\end{eqnarray} 
where the trace was performed using the basis described in eqs. 
(\ref{eq:vacuo} 
- \ref{eq:cascomut}) and $\alpha_{m+1} = r \alpha_{m} (1 - 
\alpha_{m})$ . 
We take the simplest case where $\alpha_{0}$ and $r$ are independent 
of $q$. 

We suppose that the dispersion 
relation of the quasi-particle is given by (equivalent to the Debye 
approximation),

\begin{equation}
    \epsilon_{q} = \epsilon(q) = \gamma q \hspace{0.3cm} , 
    \label{eq:dispersao}
\end{equation} 
and we enclose the system in a large 3-dimensional volume $V$.  
Replacing, in the usual way 
(since we are considering phonons with a spectrum different 
from the harmonic oscillator one), 
the sum over particles by an integral over a 
$q$-space, 

\begin{equation}
    \sum_{q}  \rightarrow \frac{V}{(2\pi)^3} \int d^3 q 
\hspace{0.3cm} ,
    \label{eq:somaint}
\end{equation} 
we obtain, for the logarithm of the partition function, after 
integrating over the angular variables,

\begin{equation}
    lnZ = \frac{V}{2 \pi^2} \int_{0}^{q_{M}} dq q^2 \ln \left( 
\sum_{m=0}^{\infty} 
    \exp(-\beta \gamma q \alpha_{m}) \right) \hspace{0.3cm} ,
    \label{eq:logz}
\end{equation} 
where this integral is evaluated over a finite q-range corresponding 
to a finite number of oscillators, and $q_{M}$ is the larger 
possible  
number $q$.  The mean energy of the solid becomes after
defining a new variable 
$\eta = \beta \gamma q$, $E_{0} \equiv \gamma q_{M}$,  
$T_{0} = E_{0}/k_{B}$ and $A \equiv \frac{V q_{M}^3}{2 \pi^2}$ 

\begin{equation}
    E = - \frac{\partial lnZ}{\partial \beta} = A 
E_{0}(\frac{T}{T_{0}})^4 \int_{0}^{T_{0}/T} 
    d\eta \eta^3 
\frac{\sum_{m=0}^{\infty} \alpha_{m} \exp(-\eta 
\alpha_{m})}{\sum_{m=0}^{\infty} \exp(-\eta \alpha_{m})} 
\hspace{0.3cm} .
    \label{eq:energia2}
\end{equation}

Let us study the integrand of eq. (\ref{eq:energia2}) .  The sum is 
performed 
over the integer $m$ that corresponds to the level of iteration of 
the 
logistic map since $\alpha_{m}$ is given by this map.  In what 
follows 
we shall consider two cases:  an example of the two-cycle and another 
one 
corresponding to the chaotic region of the logistic map.  

At a given approximation, in the two-cycle region of the logistic map 
\textbf{($3 < r < 3.449489\ldots$)}, the 
iteration runs over transient states before reaching the asymptotic 
two levels, which are infinitely degenerated.  
%
%
Clearly, when the degeneracy $g$ of the two levels goes to infinity 
the contribution of the transient states 
disappears and only the contribution of the 
states related to the asymptotic levels remains. The measure is 
concentrated 
on the two asymptotic levels.  The effective expression for the energy 
in the infinite $g$ limit is given by: 

\begin{equation}
     E = A E_{0} (\frac{T}{T_{0}})^4  
     \int_{0}^{T_{0}/T}d\eta \eta^3 
    \frac{ 
    \alpha^{-} \exp(- \eta \alpha^{-}) + \alpha^{+} \exp(- \eta 
\alpha^{+})}
    { \exp(- \eta \alpha^{-}) +  \exp(- \eta \alpha^{+})} 
\hspace{0.1cm} . 
    \label{eq:enciclo2ef}
\end{equation}

\begin{figure}[hbt]
\centerline{\epsfig{file=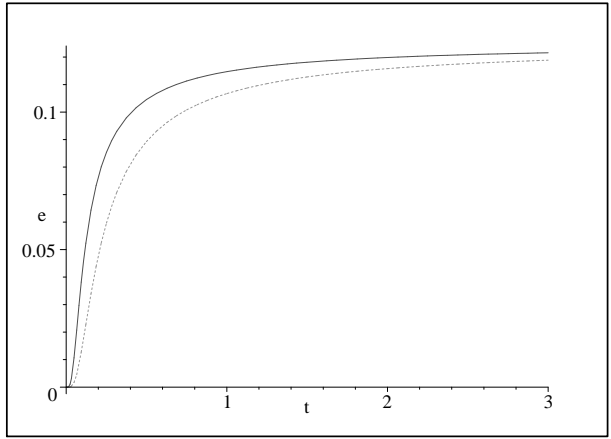,width=8cm,angle=270}} 
\caption{{\small Two cycle
energy versus temperature. Continuous line, $r=3.1$. 
Broken line, $r=3.35$ }  } 
\label{energiac2} 
\end{figure}

\begin{figure}[hbt]
\centerline{\epsfig{file = 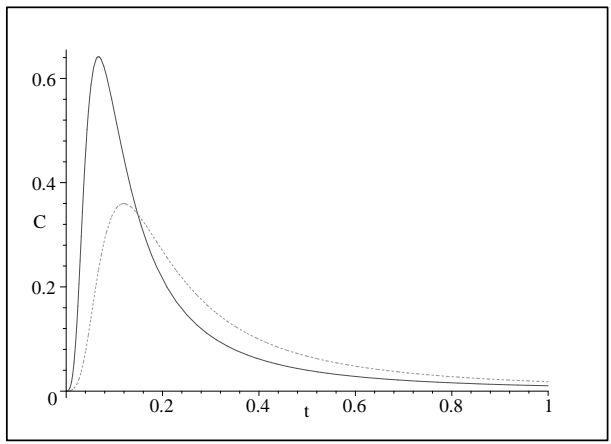,width=8cm,angle=-90}} 
\caption{{\small Specific heat for a two cycle. Continuous line, 
$r=3.1$; broken line, $r=3.35$.}} 
\label{cvc2} 
\end{figure}

For the specific heat at constant volume we have:

\begin{equation}
    C_{V} = \left(\frac{\partial E}{\partial T} \right)_{V} = 
    A k_{B} (\frac{T}{T_{0}})^3 
    \left[ 4 \int_{0}^{T_{0}/T}d\eta \eta^3 f(\eta) - 
    (\frac{T_{0}}{T})^4 f(T_{0}/T) \right] 
    \hspace{0.2cm} ,
    \label{eq:cvciclo2}
\end{equation}
where 

\begin{equation}
    f(\eta) = \frac{\alpha^{-} \exp(- \eta \alpha^{-}) + 
    \alpha^{+} \exp(- \eta \alpha^{+})}
    { \exp(- \eta \alpha^{-}) +  \exp(- \eta \alpha^{+})} 
\hspace{0.3cm} .
    \label{eq:feta}
\end{equation}
In figure \ref{energiac2} we display  $e \equiv E / A E_{0}$ times 
$t \equiv T/T_{0}$; 
in figure \ref{cvc2} we show  $C \equiv C_{V} / C_{0}$ times $t$ with 
$C_{0} \equiv A k_{B} $ .  These are typical graphics for two-level 
systems since after the transient states what remains is the 
two-cycle 
situation. For higher-cycle regions of the logistic map we shall 
have the typical behavior of a system with a finite number of 
levels.  

If we calculate the entropy from eq. (\ref{eq:logz}) we see that it 
diverges, since the degeneracy factor $g$ goes to infinity. The 
renormalized 
entropy  $S_{R} \equiv (S/k - (A/3) \log g )/A $ can be calculated and 
expressed as: 

\begin{eqnarray}
    S_{R}  & = & (\frac{T}{T_{0}})^3 \left[ \int_{0}^{T_{0}/T} d\eta \eta^2 
    \log\left( \exp(-\eta \alpha^{+}) + \exp(-\eta \alpha^-) \right) + \right.
     \nonumber  \\   
    & &  \left. \int_{0}^{T_{0}/T} d\eta \eta^3 f(\eta)  \right] \hspace{0.3cm} .
    \label{eq:s2}  
\end{eqnarray}

\begin{figure}[h]
\centerline{\epsfig{file = 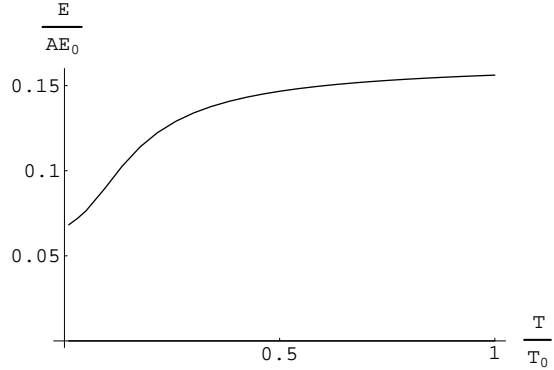,width=8cm}} 
\caption{{\small Energy versus temperature for a chaotic spectrum.} }
\label{energiabanda} 
\end{figure}

\begin{figure}[h]
\centerline{\epsfig{file = 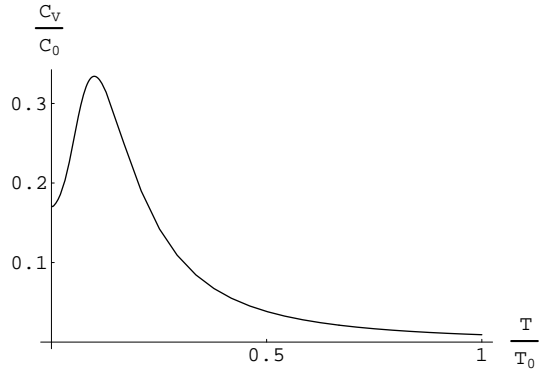,width=8cm}} 
\caption{{\small Specific heat versus temperature for the chaotic spectrum.} } 
\label{cvbanda} 
\end{figure}

\noindent 
More interesting is the behavior  of the system we are analysing for 
the chaotic region.  In this case we have as before transient 
states with the 
difference that instead of having a finite number of asymptotic 
levels we 
have a continuum of levels similar to the classical continuum levels 
in a 
classical system.  Thus, after dropping the transient states, 
as the measure is concentrated on the chaotic region, the system 
is better described by a density function that represents the number 
of hits 
of the logistic map in the interval $[0,1]$ .  In this case the mean 
energy is given by 

\begin{equation}
    E = A E_{0} (\frac{T}{T_{0}})^4 
    \int_{0}^{T_{0}/T} d\eta \eta^3 F(\eta) \hspace{0.3cm} ,
    \label{eq:enbanda}
\end{equation} 
with 

\begin{equation}
    F(\eta) = \frac{\int_{0}^{1}d\xi \xi H(\xi) \exp(-\eta \xi)}
    {\int_{0}^{1}d\xi H(\xi) \exp(-\eta \xi)} ,
    \label{eq:Fbanda}
\end{equation} 
where  $H(\xi)$ is the density function.  The specific heat at 
constant 
volume becomes 

\begin{equation}
    C_{V} = \left(\frac{\partial E}{\partial T} \right)_{V} = 
    C_{0} (\frac{T}{T_{0}})^3 
    \left[ 4 \int_{0}^{T_{0}/T}d\eta \eta^3 F(\eta) - 
    (\frac{T_{0}}{T})^4 F(\frac{T_{0}}{T}) \right] 
    \hspace{0.2cm} .
    \label{eq:cvbanda}
\end{equation} 
In figures \ref{energiabanda} and \ref{cvbanda}  we show $E / A 
E_{0}$ 
and $C_{V}/C_{0}$ times 
$T/T_{0}$ in the chaotic region for $r=3.6785$ . 

Figures \ref{energiabanda} and 
\ref{cvbanda} exhibit the 
typical low-temperature behavior of a classical system as we had already 
anticipated, since 
we have a continuum energy level in the chaotic band.  As the 
spectrum is limited from above, the behavior of the 
specific heat, at high temperatures, for 
any value of $r$, 
is proportional to $1/T^2$ as expected from systems that present 
the Schottky anomaly. 

\begin{figure}[h]
\centerline{\epsfig{file = 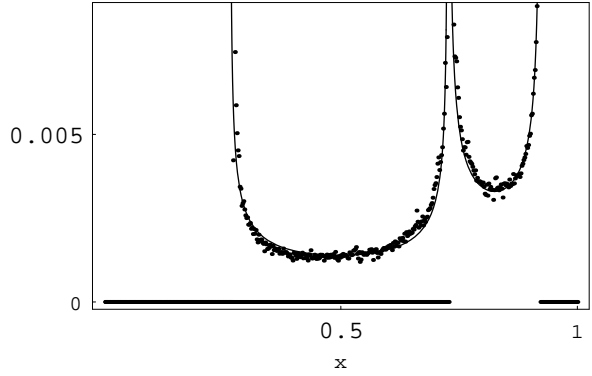,width=8cm}} 
\caption{{\small Histogram of the logistic map for $r = 3.6785$ 
and the function 
we used (full line) to calculate the thermodynamic functions.}}
\label{hist} 
\end{figure}

We also show in fig. \ref{hist}  
the normalized histogram and the density function

\begin{equation}
       H(\xi) = \left \{ 
        \begin{array}{ccc}
            \left( \; (\pi/2) \sqrt{(\xi - 0.266) (0.726 - \xi)} \; \right)^{-1} & 
	    \hspace{0.3cm} if  \hspace{0.3cm}  &  0.266 \leq \xi \leq 0.726 
	    \hspace{0.2cm} , \\ 
            \left( \; (\pi/2) \sqrt{(\xi - 0.728) (0.922 - \xi)} \; \right)^{-1} & 
	    \hspace{0.3cm} if  \hspace{0.3cm}  &  0.728 \leq \xi \leq 0.922 
	    \hspace{0.2cm} , \\
            0 &   &  otherwise \hspace{0.2cm} , 
        \end{array} 
	\right.
\end{equation}

\noindent 
we used in order to compute 
the mean energy, the specific heat and the entropy.  

After dropping the transient states, the entropy for the chaotic 
region $S_{chaos} \equiv S/S_{0}$, where $S_{0} = C_{0} = A k_{B}$, 
 can be expressed as: 

\begin{eqnarray}
    S_{chaos}  & = & (\frac{T}{T_{0}})^3 \left[ \int_{0}^{T_{0}/T} d\eta \eta^2 
    \log\left( \int_{0}^{1}d\xi H(\xi) \exp(-\eta \xi) \right) + \right. 
    \nonumber  \\   
    & &  \left. \int_{0}^{T_{0}/T} d\eta \eta^3 F(\eta)  \right] \hspace{0.3cm} .
    \label{eq:sbanda}  
\end{eqnarray}

\begin{figure}[h]
\centerline{\epsfig{file = 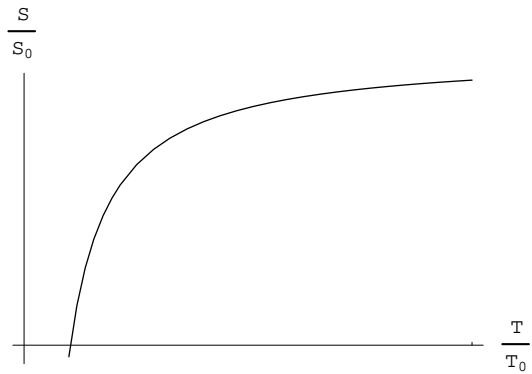,width=8cm}} 
\caption{{\small Entropy versus temperature for a chaotic spectrum.}}
\label{sbanda} 
\end{figure}

\noindent 
The entropy of the chaotic band, showed in 
fig. \ref{sbanda}, presents a curious behavior. In fact, 
its low-temperature behavior 
is typical of a classical system, with a negative divergence as 
$T \rightarrow 0$ . On the other hand, its high-temperature behavior 
is typical of a system with a limited spectrum, found 
mainly in quantum systems. The origin of this hybrid behavior 
is 
the fact that the neighbor levels in the chaotic band have no minimum 
distance 
among them, since they are dense inside the chaotic band.  This 
is somewhat equivalent of taking $\hbar \rightarrow 0$ limit, thus 
reobtaining 
classical low-temperature behaviors.  Note that in this case this 
limit is not imposed, but it is intrinsic of the system since  
the commutation relations are always different from zero.

\section{Conclusion}\label{sec.conclusao}

We construct, based on an algebra developed in \cite{marco}, a Hamiltonian 
of a quasi-particle that presents an energy spectrum whose energy levels 
are generated by the logistic map.  Depending on the parameter $r$ of the 
logistic map, the energy levels can be finite (corresponding to cycles of 
the logistic map) or chaotic (corresponding to the chaotic bands of the 
map).  We study the thermodynamic properties of a Debye-like solid constituted 
by these elementary quasi-particles and we exhibit the behavior of some 
thermodynamical quantities like internal energy, specific heat and 
entropy.  These functions, associated to the chaotic 
spectrum, present a mixed aspect, with both classical and quantum-like 
typical behaviors.  This is consequence of the fact that the spectrum in the 
chaotic region is continuous, similar to the spectra of classical systems, 
with no 
separation between neighboring levels.  On the other hand, 
the thermodynamic quantities 
related to the cycles are equivalent to systems with a finite number of 
energy-levels. 

It is interesting to note that the algebraic formalism developed in 
section \ref{sec.algebra} works consistently for a large class of maps 
$J_{i+1} = f(J_{i})$ .  Of course, changing eqs. 
(\ref{eq:logistic}, \ref{eq:logistic2}) implies  a different representation 
theory of the algebra (\ref{eq:j+}-\ref{eq:j01}) and a different physical Hamiltonian.

A classification of the analytical functions $f$ under a stability theory
would lead us to determine the different Hamiltonians associated to the
different kinds of attractors of the map $f$. A systematic study of different 
non-trivial relations $J_{i+1} = f(J_{i})$ and their consequences on the 
Hamiltonian spectra is under study.

\vspace{1cm}

We acknowledge interesting remarks from C. Tsallis. We also 
thank financial support by CNPq and Pronex/MCT.

\end{document}